\documentclass[twocolumn]{aastex63}

\usepackage{graphicx}	
\usepackage{amsmath}	
\usepackage{amssymb}	
\usepackage{xspace}

\newcommand{\Ha}{H$\alpha$\xspace}

\newcommand{\hdue}{$\rm H_{2}$\xspace}
\newcommand{\Msun}{$\rm M_\odot$\xspace}

\newcommand{\aco}{$\alpha_{CO}$\xspace}
\received{}
\revised{}
\accepted{\today}
\submitjournal{ApJL}

\shorttitle{H2 in jellyfish galaxies}
\shortauthors{Moretti et al.}
\graphicspath{{./}{figures/}}

\begin{document}
\title{The high molecular gas content, and the efficient conversion of neutral into molecular gas, in jellyfish galaxies}

\correspondingauthor{Alessia Moretti}
\email{alessia.moretti@inaf.it}

\author[0000-0002-1688-482X]{Alessia Moretti}
\affiliation{INAF-Padova Astronomical Observatory, Vicolo dell'Osservatorio 5, I-35122 Padova, Italy}

\author[0000-0001-9143-6026]{Rosita Paladino}
\affiliation{INAF-Istituto di Radioastronomia, via P. Gobetti 101, I-40129 Bologna, Italy}

\author[0000-0001-8751-8360]{Bianca M. Poggianti}
\affiliation{INAF-Padova Astronomical Observatory, Vicolo dell'Osservatorio 5, I-35122 Padova, Italy}

\author[0000-0001-5965-252X]{Paolo Serra}
\affiliation{INAF-Cagliari Astronomical Observatory, Via della Scienza 5, I-09047 Selargius (CA), Italy}

\author[0000-0003-0231-3249]{Mpati Ramatsoku}
\affiliation{Department of Physics and Electronics, Rhodes University, PO Box 94, Makhanda, 6140, South Africa}
\affiliation{South African Radio Astronomy Observatory, 2 Fir Street, Black River Park, Observatory, Cape Town, 7405, South Africa}
\affiliation{INAF-Cagliari Astronomical Observatory, Via della Scienza 5, I-09047 Selargius (CA), Italy}

\author[0000-0001-9575-331X]{Andrea Franchetto}
\affiliation{Dipartimento di Fisica e Astronomia “Galileo Galilei”, Universit\'a di Padova, vicolo dell’Osservatorio 3, IT-35122, Padova, Italy}
\affiliation{INAF-Padova Astronomical Observatory, Vicolo dell'Osservatorio 5, I-35122 Padova, Italy}

\author{Tirna Deb}
\affiliation{Kapteyn Astronomical Institute, University of Groningen, Postbus 800, NL-9700 AV Groningen, the Netherlands}

\author[0000-0002-7296-9780]{Marco Gullieuszik}
\affiliation{INAF-Padova Astronomical Observatory, Vicolo dell'Osservatorio 5, I-35122 Padova, Italy}

\author[0000-0002-8238-9210]{Neven Tomi\v{c}i\'{c}}
\affiliation{INAF-Padova Astronomical Observatory, Vicolo dell'Osservatorio 5, I-35122 Padova, Italy}

\author[0000-0003-2589-762X]{Matilde Mingozzi}
\affiliation{INAF-Padova Astronomical Observatory, Vicolo dell'Osservatorio 5, I-35122 Padova, Italy}

\author[0000-0003-0980-1499]{Benedetta Vulcani}
\affiliation{INAF-Padova Astronomical Observatory, Vicolo dell'Osservatorio 5, I-35122 Padova, Italy}

\author[0000-0002-3585-866X]{Mario Radovich}
\affiliation{INAF-Padova Astronomical Observatory, Vicolo dell'Osservatorio 5, I-35122 Padova, Italy}

\author[0000-0002-4158-6496]{Daniela Bettoni}
\affiliation{INAF-Padova Astronomical Observatory, Vicolo dell'Osservatorio 5, I-35122 Padova, Italy}

\author[0000-0002-7042-1965]{Jacopo Fritz}
\affiliation{Instituto de Radioastronomia y Astrofisica, UNAM, Campus Morelia, AP 3-72, CP 58089, Mexico}

\begin{abstract}
In the disks of four jellyfish galaxies from the GASP sample at redshift $\sim 0.05$ we detect molecular gas masses systematically higher than in field galaxies.
These galaxies are being stripped of their gas by ram pressure from
the intra cluster medium and are, in general, forming stars at high rate with respect to non-stripped galaxies of similar stellar masses.
We find that, unless giant molecular clouds in the disk are unbound by ram pressure leading to exceptionally high CO--to--$\rm H_2$ conversion factors, these galaxies have a molecular gas content 4-5 times higher than normal galaxies of similar masses, and molecular gas depletion times ranging from $\sim$1 to 9 Gyr, corresponding to generally very low star formation efficiencies.
The molecular gas mass within the disk is a factor between  4 and $\sim$100
times higher than the neutral gas mass, as opposed to the disks of normal spirals that contain similar amounts of molecular and neutral gas.
Intriguingly, the molecular plus neutral total amount of gas is similar to that in normal spiral galaxies of similar stellar mass. These results strongly suggest that ram pressure in disks of galaxies during the jellyfish phase leads to a very efficient conversion of HI into \hdue.
\end{abstract}


\section{Introduction} \label{sec:intro}
What ultimately regulates galaxy evolution is the availability of gas prone to star formation, and its efficiency in forming new stars \citep{Kennicutt1998,Schmidt1959}.
Therefore, a lot of effort has been put in the last years in building statistically significant samples of galaxies with an observational coverage able to map all the gas phases, with the aim of discovering the underlying scaling relations that correlate the neutral and molecular gas content with the stellar mass, if any \citep{Saintonge2011,Saintonge2011b,Catinella+2018,Corbelli+2012,Cortese+2011}.
More recent results also include the dust contribution \citep{Casasola+2020}.

Scaling relations have been derived, in fact, for the HI content of nearby galaxies \citep{Bigiel2008,Bigiel2010} while only small samples exist at higher redshifts \citep{chiles,Cortese+2017,Catinella+2018}, suggesting that the gas in the molecular phase was the predominant fraction at early epochs. 

As for the cold gas content, the largest statistical sample for which an homogeneous set of data has been collected so far is the COLDGASS \citep{Saintonge2011} sample together with its low mass extension xCOLDGASS \citep{Saintonge2017}.
Galaxies in the xCOLDGASS sample are mass--selected at z=0.01-0.05, without any {\it a priori} bias on the IR/UV fluxes, and cover a range in stellar mass between $10^9$ and $10^{11.5} M_{\odot}$. 
While this sample gives a wonderful insight into the gas properties of galaxies in general, it has not been designed to cover galaxies in different environmental conditions, i.e. it does not distinguish between galaxy properties in clusters and those in the field, which are expected to behave differently.
While early works on the Virgo and Coma clusters \citep{Kenney+1989, Boselli+1997} have revealed no significant molecular gas deficiency in cluster galaxies (including the most HI deficient), more recent studies that make use of larger samples and better resolution find, instead, that HI gas stripped galaxies have also a lower \hdue content, albeit in lower proportion \citep{Boselli+2014}.
Similar results are also given by \citet{Corbelli+2012}, that studies 35 metal rich spiral galaxies that are part of the Herschel Virgo Cluster Survey. 
These galaxies have a well determined HI deficiency parameter, which has been found to anti-correlate with the molecular gas fraction, i.e. the ratio between molecular gas mass and stellar mass. Indeed, the molecular gas fraction has been found to decrease as the HI deficiency increases, while the ratio between the molecular and the total gas increases. This has been interpreted as due to the fact that both neutral and molecular gas are stripped, but the former is stripped more easily than the latter.

Single dish and, more recently, interferometric ALMA data have confirmed a normal molecular gas content in the disk of ESO137-001, a jellyfish galaxy in the nearby Norma cluster \citep{Jachym2014,Jachym+2019}, albeit confined in a very small central region (with $\sim 1.5$ kpc radius), with a similar amount of molecular gas detected along the stripped tail.
The D100 galaxy close to the Coma cluster center, instead, shows a higher than expected molecular gas fraction within the central $\sim 2$ kpc radius \citep{Jachym2017}, and a very \hdue rich gas tail, as traced by the CO emission.

A clear view on the molecular gas content of cluster galaxies subject to ram-pressure stripping is still missing, and we can now start to cast light on the subject by using the multiwavelength dataset collected by the GASP sample.
The GASP survey\footnote{https://web.oapd.inaf.it/gasp/} \citep{gaspI} has started exploring the effects of environmental interactions on nearby ($z\sim0.05$) cluster/group galaxies thanks to a dedicated VLT MUSE Large Program (GAs Stripping Phenomena in galaxies with MUSE, P. I. B. Poggianti) mainly tracing the ionized gas, but complementary datasets at different wavelengths are being collected and have started offering a clear view on all the connected gas phases \citep[see also][]{gaspXVII,George+2018,Deb+2020,GASPXXVIII,GASPXXIII}.
In particular, in \cite{gaspX} we observed with the APEX telescope 4 GASP galaxies detecting molecular gas both in the galaxy disks and in the ionized gas tails.

We then started an observational campaign devoted at measuring the molecular gas content of these galaxies using ALMA interferometric data at $\sim$1 kpc resolution. 
We have described the observations and the data analysis in \citet{GASPXXII}, where we have shown the results for the JW100 galaxy.
Surprisingly, in this galaxy we have found an anomalously large content of molecular gas, even excluding the new molecular gas possibly born in the tail from the stripped neutral gas.
Assuming the standard conversion factor, we tentatively concluded that the star formation efficiency (SFE), i.e. the Star Formation Rate (SFR) surface density divided by the \hdue mass density, on scales of 1 kpc shows a gradient moving from the central part of the disk toward the stripped tail, and the corresponding depletion time is always longer than the typical $\tau_{dep}$ of nearby disk galaxies (i. e. $\sim 1-2$ Gyr, \citealt{Bigiel2011,Leroy2013}), increasing from the disk to the tail.

In this paper we analyze the ALMA Band 3 data of the four GASP galaxies observed in cycle 5 to understand whether their molecular gas content is following the scaling relations of normal galaxies, or if instead they are depleted/enriched, in molecular gas. 

All the GASP galaxies here analyzed are operatively defined as jellyfishes (as they possess ionized gas tails whose length is comparable to the galaxy disk diameter, \citealt{poggianti2017}), and they also have a central AGN.
Stellar masses and redshifts are given in \citet{gaspX}, while results from MUSE data are described in \citet{gaspI,gaspII,gaspIV,GASPXXIII,GASPXXII}.

Throughout this paper we will make use of the standard cosmology $H_0 = 70 \, \rm km \, s^{-1} \, Mpc^{-1}$, ${\Omega}_M=0.3$ and ${\Omega}_{\Lambda}=0.7$.
As in the other GASP papers, our stellar masses are calculated adopting a \cite{Chabrier2003} Initial Mass Function (IMF).

\section{Data and analysis}

Observations of the CO(1-0) emission (rest frequency 115.271 GHz) of our four galaxies, namely JO201, JO204, JO206 and JW100, have been obtained with ALMA during Cycle 5 (project 2017.1.00496.S). 
The spectral configuration used yields a velocity resolution of 3.1 km/s, which has been smoothed to 20 km/s in the final datacubes used for the following analysis.
In order to cover with homogeneous sensitivity the entire area observed with MUSE, including disk and tails, mosaics have been necessary.
The actual configurations used for the observations allowed us to sample scales
up to the maximum recoverable scale (MRS), reported in Tab.\ref{tab:data} for each galaxy.

The data have been calibrated and imaged using the CASA software (version 5.4.0-7; McMullin 2007), as described in \citet{GASPXXII}. The details of the obtained line images are reported in Tab. \ref{tab:data}

\begin{table}
\begin{center}
\caption{Properties of CO(1-0) line images: observed frequency, 
synthesized beam ($\theta_{maj}$, $\theta_{min}$, and PA), rms and maximum recoverable scale (MRS; eq. 7.7 in the ALMA Technical Handbook).
}\label{tab:data}
\begin{tabular}{|l|c|c|c|c|c|c|}
\hline
Galaxy & $\nu_{\rm obs}$ &$\theta_{maj}$ & $\theta_{min}$  & PA   &rms      & MRS \\ 
       & GHz&  $^{\,\prime\prime}$ & $^{\,\prime\prime}$          & deg  & mJy/b  &  $^{\,\prime\prime}$   \\ 
\hline
\hline
JO201  &110.307 &1.99 & 1.57  &  -84.6  & 0.5     &20   \\ 
JO204  & 110.625 &1.62 & 1.36  &   81.5  & 0.5     & 20  \\ 
JO206  & 109.677 &1.60 & 1.30  &  -87.4  & 0.7     & 23 \\ 
JW100 & 108.644 &2 & 1.7 & 8.3 & 0.9 & 24 
\\
\hline\hline
\end{tabular}
\end{center}
\end{table}

From the cleaned datacubes we obtained moment-zero maps, using the {\sc SoFiA} software \citep{sofia} to construct detection masks for each ALMA datacube, as described in \citet{GASPXXII} for JW100.
Fig.\ref{fig:maps} shows the CO(1-0) moment-zero for the disks of the four galaxies analyzed here.

For each galaxy we measure both the amount of molecular gas within the stellar disk and outside it, where the CO is located in correspondence with the ionized gas tails found in \citet{gaspI,gaspII,gaspIV,GASPXXIII}.
We use here as disk definition the one given in Gullieuszik+ submitted (shown as red contour in Fig.\ref{fig:maps}), which is based on the stellar isophote at 1$\sigma$ above the sky background level measured on the undisturbed side of the galaxy and then symmetrized to exclude the contribution from the stripped tail.

Tab. \ref{tab:data} shows that we will base our results on the detection of molecular gas on scales between $\sim 1$ and $\sim 20$ kpc, thus neglecting the contribution from any gas diffuse on larger scales. 
We do not have single dish data at the CO(1-0) frequency to assess the possible flux loss, and therefore our data are, strictly speaking, lower limits.
The uncertainty on the measured fluxes is of the order of $10\%$.

\section{Results}
\begin{figure*}
    \centering
    \includegraphics[width=0.45\textwidth]{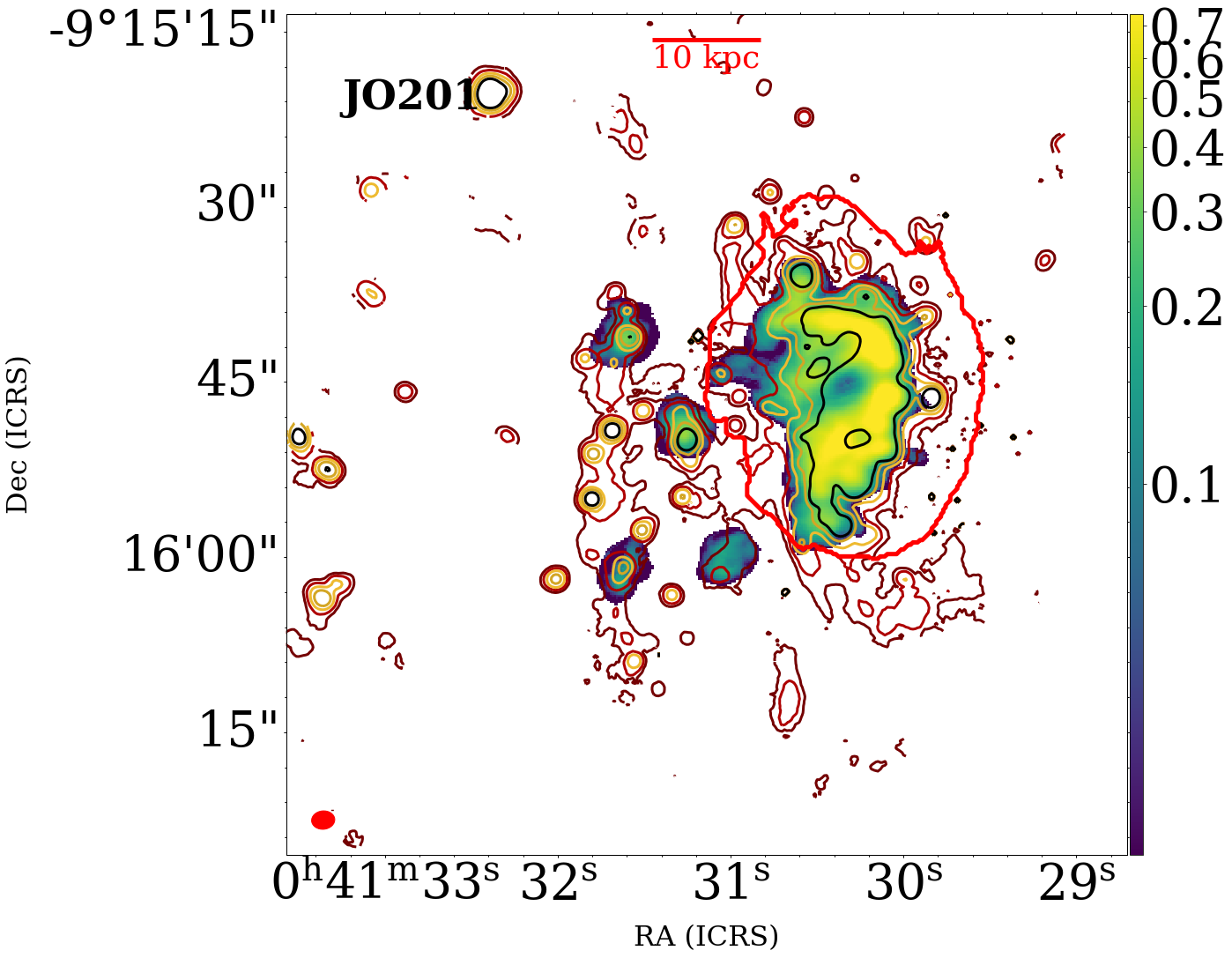}
    \includegraphics[width=0.45\textwidth]{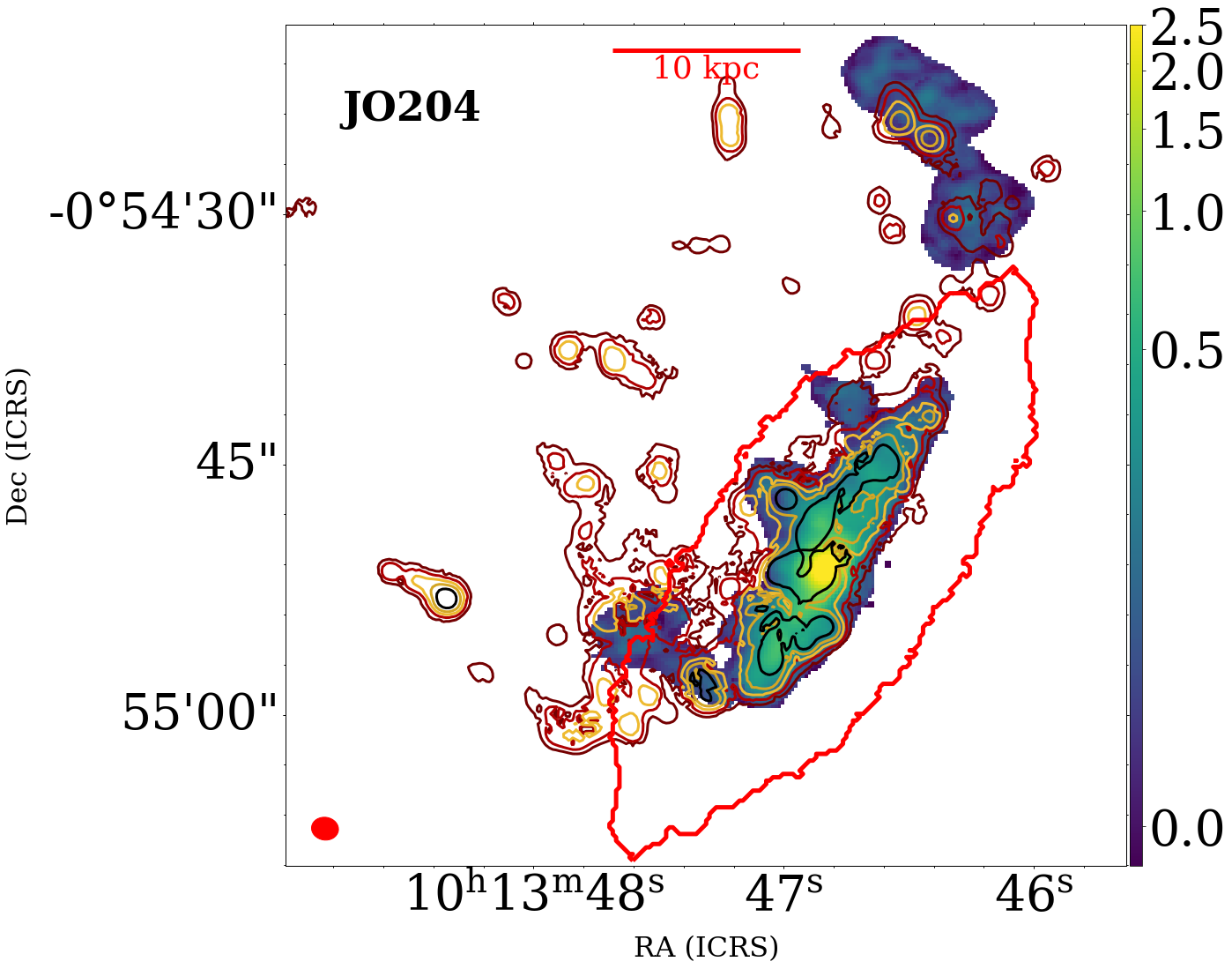}
    \includegraphics[width=0.45\textwidth]{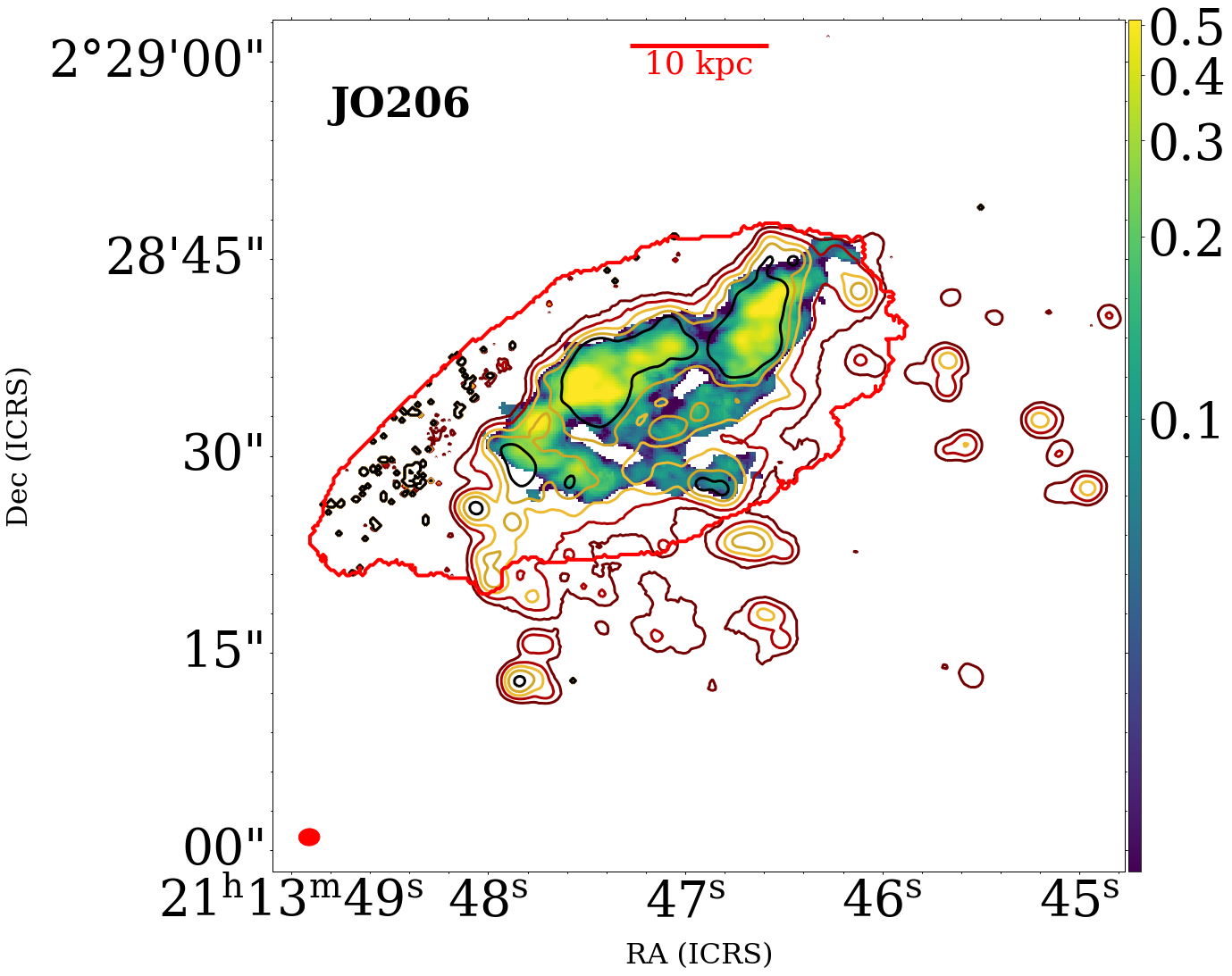}
    \includegraphics[width=0.45\textwidth]{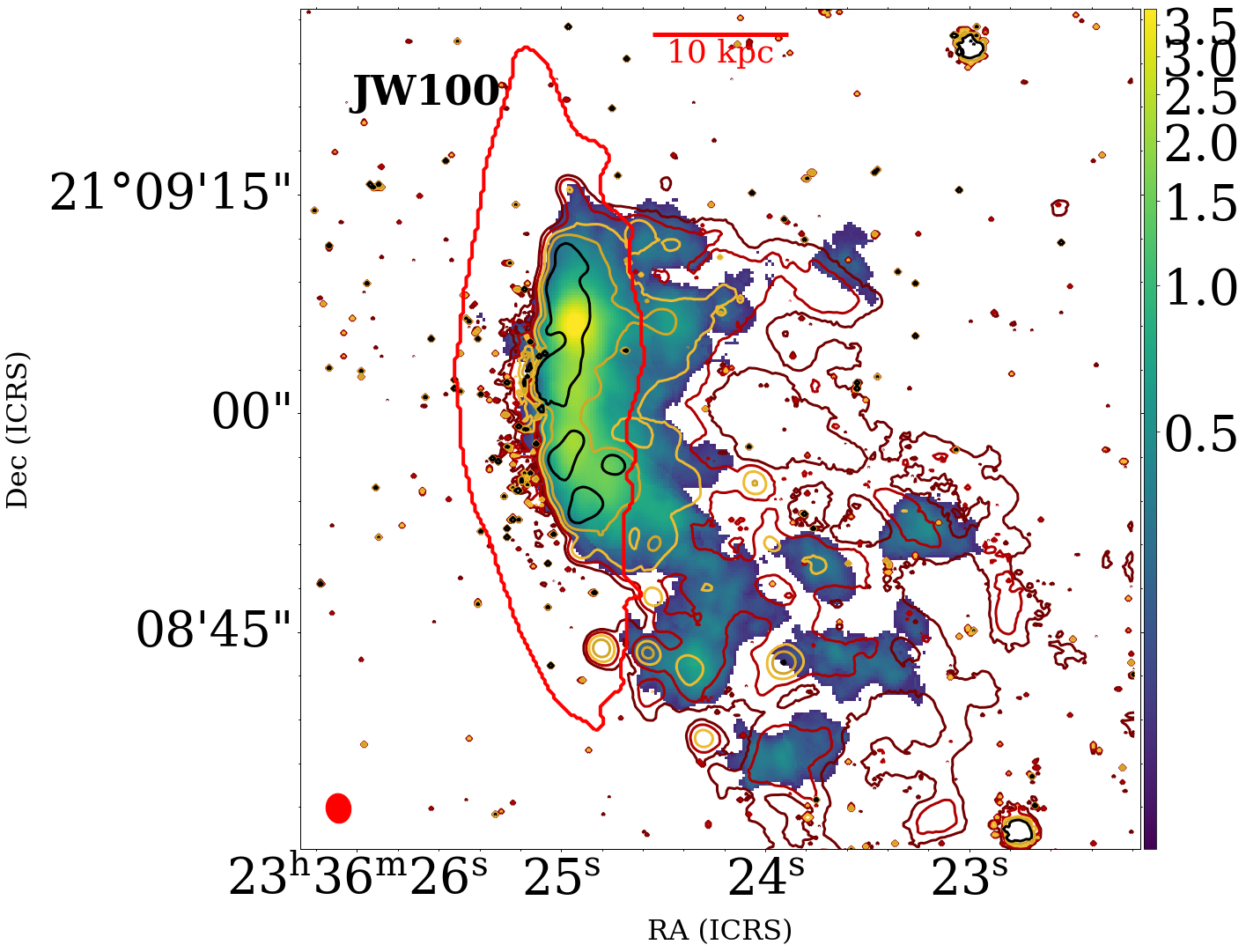}
    \caption{CO(1-0) moment-zero maps (in scale colors) in Jy/beam.km/s for JO201, JO204, JO206 and JW100. Colored contours show the \Ha emission derived from MUSE data at $2\times10^{-17}$,  $4\times10^{-17}$,  $8\times10^{-17}$,  $1.6\times10^{-16}$ and  $3.2\times10^{-15}$ erg/cm$^2$/s/arcsec$^2$, while the red contour delimits the galaxy  stellar disk derived from the MUSE data (see text for details). The scale in kpc and the beam size are also shown in red within each panel.}
    \label{fig:maps}
\end{figure*}

As shown in Fig. \ref{fig:maps}, the molecular gas distribution in the galaxy disk is quite consistent with the \Ha emission (colored contours) derived from MUSE data \citep{gaspII,gaspI,gaspIV,GASPXXIII}.

For each galaxy we have used the CO(1-0) emission to derive 
the CO flux and the \hdue mass using the following equation from \citealt{WatsonKoda2016}:
\begin{equation}
\label{eqn:wk_co10}
\left(\frac{M_{\rm H_2}}{M_{\odot}}\right) = 
1.1 \times 10^4 \left( \frac{\alpha_{co}}{4.3}\right)
\left(\int S_{10}dv \right)
\left(D_L \right)^2
\end{equation}

where $\rm \alpha_{co}$ is the CO-to-$\rm H_2$ conversion factor expressed in
$M_{\odot} \rm (K \, km \, s^{-1} \, pc^{2})^{-1}$, $S_{10}$ is the CO flux density
in Jy and D$_L$ is the luminosity distance in Mpc.

As well known, the \hdue mass is strongly dependent on the \aco factor.
Tab. \ref{tab:masses} gives the molecular gas masses for different assumptions on \aco, as well as the SFRs measured within the disk using the star forming spaxels from \cite{vulcani+2018_sf} and the corresponding \hdue depletion times, defined as $M_{\rm H_2}/SFR$. 

\begin{figure*}
    \centering
    \includegraphics[width=0.95\textwidth]{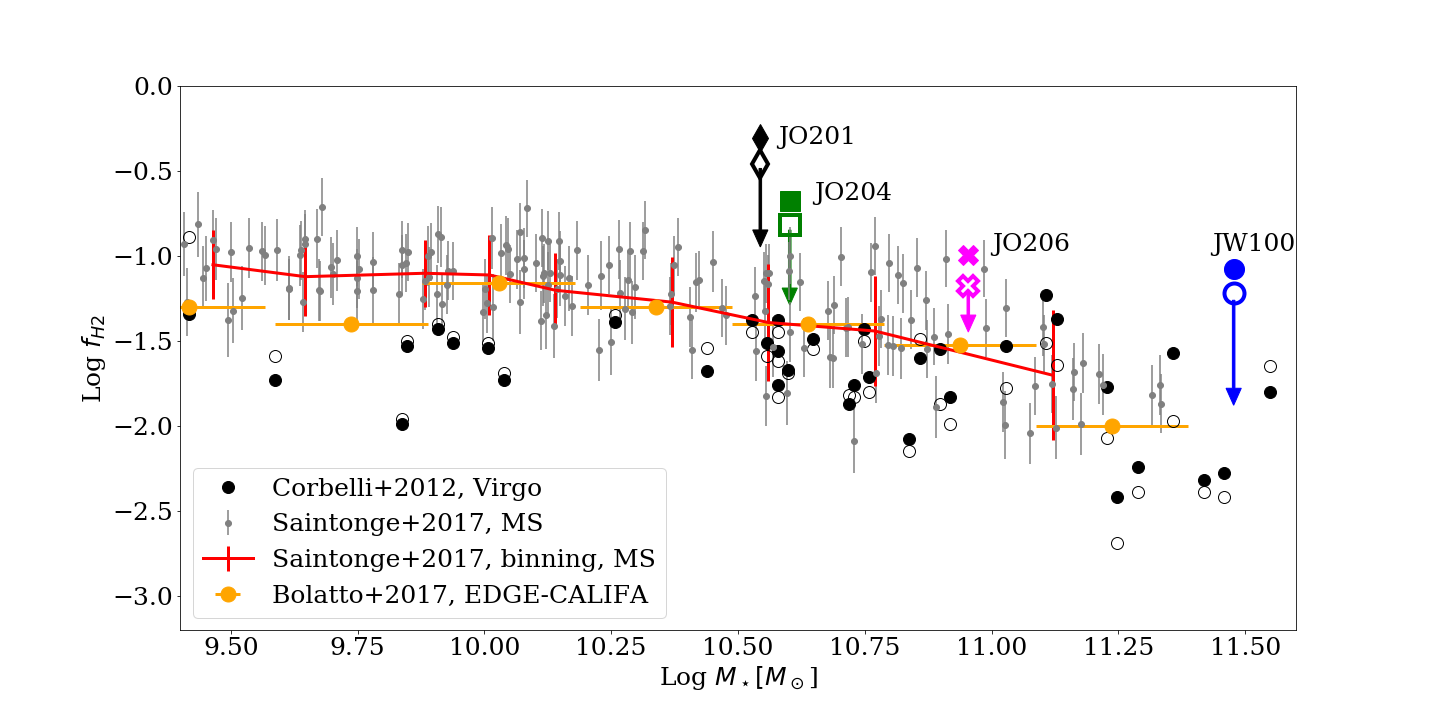}
    \caption{Total molecular gas fraction $f_{H_2}(=M_{H_2}/M_{\star}$) as a function of the stellar mass. \hdue masses have been derived using a constant \aco (filled symbols) and two different metallicity dependent \aco: the \citealt{Accurso2017} (empty symbols) and the \citealt{AMorin+2016} (arrows). The red line shows the mean scaling relation found in the xCOLDGASS survey by \citet{Saintonge2017} for field galaxies on the main sequence, and grey dots are the single measurements.
    Orange dots refer to spiral galaxies from the EDGE-CALIFA survey \citep{Bolatto+2017}.
    Black dots are HI deficient galaxies in Virgo from \citealt{Corbelli+2012}: filled symbols and empty symbols represent \hdue masses derived with constant or metallicity dependent \aco.
    }
    \label{fig:fh2}
\end{figure*}

 First, we used the Milky Way \aco that is equal to 4.3 
 $M_{\odot} \rm (K \, km \, s^{-1} \, pc^{2})^{-1}$ \citep{Bolatto2013}, including the Helium correction, which is the standard value used in the literature.
We used this value to calculate the molecular gas masses and the corresponding gas fractions with respect to the galaxy stellar mass, $f_{H_2}=M_{H_2}/M_{\star}$, shown as filled symbols in Fig.\ref{fig:fh2}.

We used the same assumption to calculate the molecular gas mass present in the tail clumps. As described in details in Moretti et al. (in prep.) that deals with the extraplanar emission, the molecular gas masses in correspondence to the tail star-forming clumps ($0.4-1.7 \times 10^9 M_{\odot}$) amounts to only a small fraction of the total stellar mass (see col. 5 in Table~\ref{tab:masses}).

As a second step, in order for our data to be comparable with the xCOLDGASS sample, we calculated the \hdue masses assuming an \aco variable with the metallicity, following the relation found for the same sample by \citet{Accurso2017}, that depends both on the galaxy metallicity and on its distance from the star formation main sequence.
This relation, though, holds only up to metallicities of 12+log(O/H)=8.8.

We used the [OIII]/[SII+] vs. [NII]/[SII+] line ratio to derive the spatially resolved gas phase metallicities \citep{Franchetto_2020} for our galaxies and derived the mean \aco within the disk. As 
our method to estimate the gas metallicity is different from the one used by \citep{Accurso2017} our galaxies might be skewed towards higher values. This does not bias the results, though, as we mostly use the asymptotic value from the \citealt{Accurso2017} relation.

As for the distance from the main sequence of the SFR-Mass relation, we assumed the average value found for jellyfish galaxies in \citealt{vulcani+2018_sf}, i.e. 0.15 dex.

Both using the Milky Way \aco (filled symbols in Fig.~\ref{fig:fh2}) and the metallicity-dependent one (empty symbols), we find that the total molecular gas fractions in our jellyfish galaxies are significantly larger than the mean values found for starforming main sequence galaxies in the xCOLDGASS sample by \citet{Saintonge2017}, shown as a reference in Fig.~\ref{fig:fh2}.  Grey dots in Fig.~\ref{fig:fh2} are the single datapoints from \citet{Saintonge2017} for galaxies within 0.4 dex from the main sequence, demonstrating indeed that our jelllyfishes lie at the upper edge of the observed distribution.
We note that the results obtained by \cite{Bolatto+2017} on a subsample of the CALIFA galaxies observed with the CARMA interferometer are perfectly in agreement with the Saintonge relation.
The Virgo cluster data by \citet{Corbelli+2012}, represented as black filled dots in Fig.~\ref{fig:fh2}, having rescaled the stellar masses to correct for the different  assumption in the IMF, lie below the mean field relations which led the authors to conclude that \hdue has been stripped from the galaxies. 

\begin{table*}
\centering
\caption{For each galaxy and \aco assumption ($^a$ from \citet{Accurso2017}, $^b$ from \citet{AMorin+2016}) we list: Stellar masses; SFRs from \Ha MUSE emission within the disks from \cite{vulcani+2018_sf}; molecular gas masses within the galaxy disk assuming different \aco and, in parenthesis, the molecular gas to stellar mass fractions in the disk; 
molecular gas to stellar mass fractions in the stripped tail;
\hdue depletion times; HI masses within the disk from \cite{gaspXVII,GASPXXVIII}, Deb+in prep.; $R_{mol}=M_{H_2}/M_{HI}$ in the disks; total gas fractions (molecular+neutral, disks+tails) and corresponding total depletion times for the different \aco.
}\label{tab:masses}
\begin{tabular}{l|l|l|lll|llll}

 $M_{\star}$& SFR & \aco &  $M_{H_2,in}$&  $M_{H_2,out}$ & $\tau_{dep,H_2}$&$M_{HI}$ & $R_{mol}$ & $M_{gas}/M_{\star}$  & $\tau_{dep,tot}$ \\
 1e9 \Msun &  \Msun/yr &  & 1e9 \Msun&  & Gyr &1e9 \Msun & & & Gyr\\
 \hline 
& & & & JO201 & & & \\
\hline 
         &            & 4.3 & 16.5[46\%] & 2\%           & 3.3 &      & 14 & 0.53 & 3.5\\
35.5     & 5$\pm$1 &3.0$^a$    & 11.5[32\%] &  & 2.3 & 1.15 & 10 & 0.39 & 2.6 \\
         &            & 1.3$^b$ & 5.0[14\%]  &     & 1   &      &  4 & 0.21 & 1.3\\
\hline
& & & & JO204 & & & \\
\hline 
   &             & 4.3 & 8.1[20\%] & 1\%    & 5.4 &   & -&-&-\\
40 & 1.5$\pm$0.3 & 3.0$^a$ & 5.7[14\%] &  & 3.9 & - & -&-&-\\
   &             & 1.4$^b$ &2.7[ 7\%] &     & 1.8 &   & -&-&-\\
\hline
& & & & JO206 & & & \\
\hline 
 &             & 4.3 & 8.7[10\%] & 0.4\%          & 1.8 &      & 12& 0.14 & 1.9\\
90 & 4.8$\pm$0.9 &2.8$^a$ & 5.6[ 6\%] &  & 1.2 & 0.7  &  8& 0.10 & 1.3\\
&             & 2.0$^b$ & 4.0[ 4\%] &       & 0.8 &      &  6& 0.08 & 1.0\\
\hline
& & & & JW100 & & & \\
\hline 
    &             & 4.3    & 23.7[ 8\%] &  0.6\%   & 9.1 &     & 132& 0.09 & 9.2\\
300 & 2.6$\pm$0.5 & 3.0$^a$& 16.5[ 5\%] &          & 6.3 & 0.2 & 92& 0.07 & 6.4\\
    &             & 0.9$^b$& 5.0[ 2\%]  &          & 1.9 &     &  28& 0.03 & 2.0\\

\end{tabular}
\end{table*}

We notice that the ram-pressure acting on the infalling galaxies could in principle unbind the already existing giant molecular clouds (GMCs) in the galaxy disk, increasing the CO/$H_2$ ratio. If this were the case, for a correct estimate of the molecular gas mass we should be using a lower \aco, more similar to that found, for example, in ULIRGs \citep[see][]{Bolatto2013,Sandstrom+2013,Israel+2015} where the underlying cold gas distribution is more diffuse. 
To mimic this effect, we also used a second metallicity-dependent formulation of \aco \citep{AMorin+2016} (which gives results compatible with the  studies 
by \citet{Schruba+2012,Genzel+2012} for higher redshift galaxies) where this factor reaches values that go from 0.9 (in JW100) to 2.0 (in JO206).
The corresponding molecular gas masses are also given in Tab.~\ref{tab:masses}.

With these assumptions we find the molecular gas fractions shown in Fig.~\ref{fig:fh2}  as arrows, which are closer to the literature relations.
We caution, however, that these extremely low conversion factors have been extrapolated using relations meant to include primarily low metallicity galaxies.  
The behaviour of the \aco in the high metallicity range is still debated, as some authors do not find significant deviations from the MW value \citep{Wolfire+2010,Sandstrom+2013}.

We conclude that the molecular gas fractions we derive for our jellyfish galaxies are much higher than in both field galaxies and in the Corbelli et al. Virgo cluster galaxies. 
Using a constant MW--like \aco, the molecular gas fractions at given stellar mass are higher than the mean values in xCOLDGASS galaxies by an average factor of $\sim 5$, while using the \citet{Accurso2017} relation by an average factor of $\sim 4$, with molecular gas fractions ranging between $\sim$8\% and $\sim$50\%.
The derived $H_2$ masses are extremely high, ranging 
between 8 and 24 $\times 10^9 \, M_{\odot}$.
Using a second metallicity-dependent \aco (plausible if the molecular gas were much more diffuse than usual due to the ram pressure actually disrupting GMCs, but highly uncertain at the high metallicities of our galaxies) yields molecular gas fractions
closer to the literature values for non-stripped galaxies, though still well above the mean for field galaxies and still higher than those observed in Virgo spirals.
Thus, unless the \aco is close to the value mostly found in ULIRGs (in the sense that most of the molecular gas is diffuse), jellyfish galaxies have huge reservoirs of $H_2$.

The standard molecular depletion time in galaxy disks, resolved on $\sim 1$ kpc scale, is $\sim 2$ Gyr \citep{Bigiel2008,Leroy2008}.
The \hdue depletion timescales of our jellyfishes are significantly longer if we consider MW-like \aco, except for JO206 (\ref{tab:masses}). 
When using the \citet{Accurso2017} scaling relation, the timescales for JW100 and JO204 are still longer than normal galaxies, while the other two are more consistent with the literature values (JO201 and JO206).
Assuming a very low \aco over the entire extent of the disk leads to timescales generally shorter than in normal galaxies.

\subsection{$H_2/HI$ mass ratio, total gas mass fractions and depletion times}
Having established that molecular gas is abundantly present in jellyfish galaxies, we now proceed to evaluate the proportion of the different gas phases within the galaxy disk to better characterize the star-forming cycle.
Three of our galaxies \citep{gaspXVII,GASPXXVIII} Deb+, in prep. also possess an estimate of the HI gas content (preliminary from MeerKat data for JW100), while for the fourth one (JO204) this has been hampered by HI absorption due to the presence of a continuum source at the center of the galaxy \citep{Deb+2020}.
We note that within the disk the neutral and the molecular gas have similar distributions, at the observed spatial scales. For the JO206 galaxy we had to recalculate the mass within the disk using the disk definition adopted here.

 \begin{figure*}
    \centering
    \includegraphics[width=0.95\textwidth]{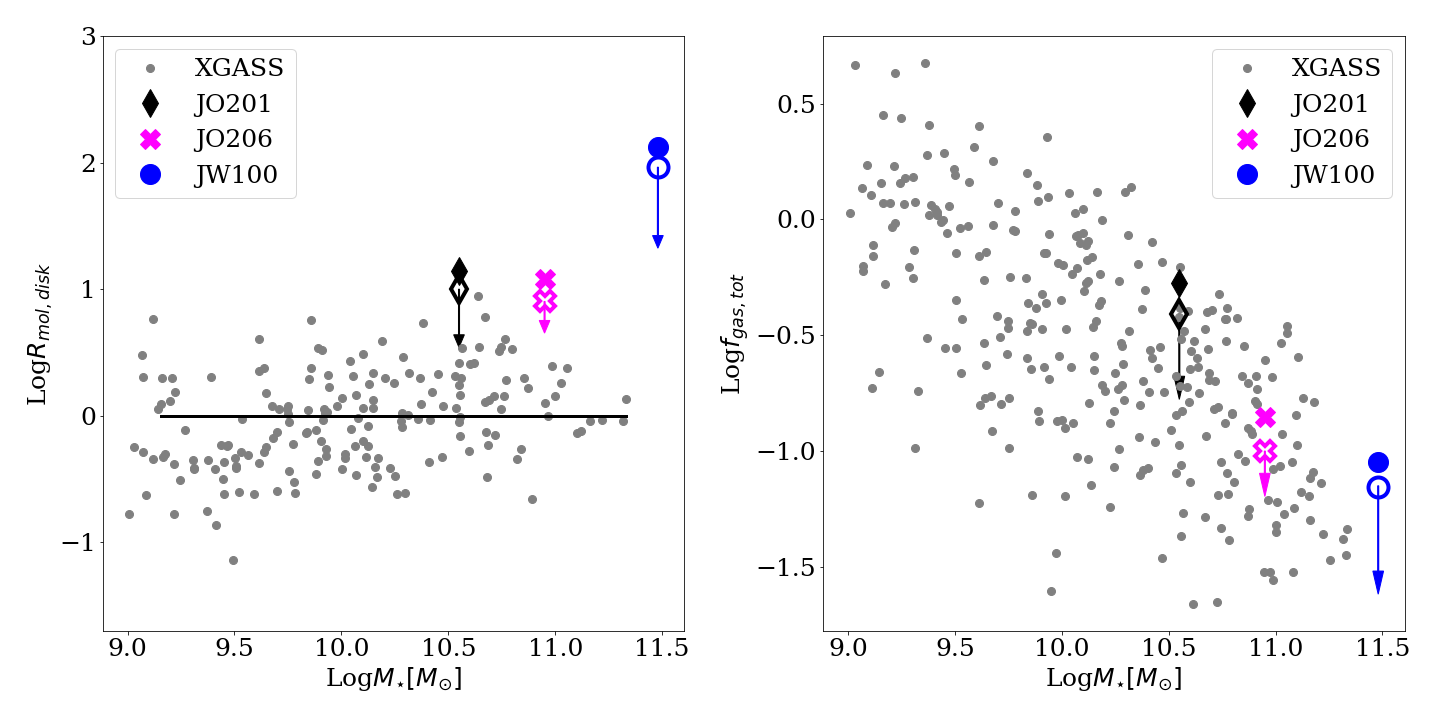}
    \caption{Molecular gas fraction  ($R_{mol,disk}=M_{H_2,disk}/M_{HI,disk}$) in the disk (left) and total gas fraction  ($f_{gas}=(M_{H_2,tot}+M_{HI,tot})/M_{\star}$) (right) for JO201, JO206 and JW100 compared with the XGASS sample as a function of the stellar mass from \citealt{Wang+2020,Catinella+2018}. The black horizontal line shows the average $R_{mol}$ of the sample.
    }
    \label{fig:scaling_rel}
\end{figure*}

In Fig.~\ref{fig:scaling_rel} we show the disk molecular gas fraction $R_{mol}=M_{H_2}/M_{HI}$ of these three galaxies, compared with the values in spiral disks derived from the xGASS sample assuming a radial fit to the HI mass distribution \citep{Wang+2020}. 
$R_{mol}$ in the disk of our galaxies turns out to be very high (ranging from 4 to 
$\sim 100$
), while the average expected value in galaxy disks is $R_{mol}=1$ (black horizontal line in Fig.~\ref{fig:scaling_rel}).
Our data are clearly significantly offset compared to normal galaxies with the same stellar mass.
This result persists for any assumption on \aco.

Since $R_{mol}$ can be considered proportional to the ratio between the typical GMC lifetime over the conversion time between the neutral and the molecular phase \citep{Leroy2008}, our results strongly suggest that the conversion of HI into molecular gas is very efficient in jellyfish galaxies. 
We note that similar trends have been found also in interacting galaxies \citep{Casasola+2004}.

Strikingly, summing up the molecular and neutral gas masses and considering the total (disk and tail) ratio of gas and stellar masses, we obtain
gas mass fractions (shown in the right panel of Fig.~\ref{fig:scaling_rel})
that are within the observed range of normal galaxies. In other words, the total gas (molecular+neutral) associated to our jellyfish galaxies (considering both what is left in the disk and what is in the tail)  is normal for their stellar mass.
These results do not change if we restrict these comparisons to normal galaxies within the range of stellar mass surface density similar to ours (that goes from 8 to 9 $M_{\odot} kpc^{-2}$).

Finally, considering depletion times, we note that in \citealt{GASPXXVIII}, for JO201 and JO206, we found that HI depletion times are much shorter than the average value found in galaxy disks, apparently implying a very efficient star formation (as far as HI is concerned) and/or a low HI content for their SFR. 
Evaluating the global (HI + \hdue) depletion time for the three galaxies where we have both measurements, it turns out that within the galaxy disk the estimated depletion time is in agreement with literature values \citep{Leroy2008} only for the JO206 galaxy, while 
JO201 and JW100 have unusually long depletion times, unless we assume
extremely low values for the \aco.

\section{Conclusions}
Our analysis of the molecular gas content of four GASP ram-pressure stripped galaxies measured on 1 kpc scale with ALMA reveals a very high molecular gas fraction with respect to both isolated/field galaxies and Virgo cluster galaxies with a similar stellar mass.

Our results are dependent on the assumptions on the \aco, but 
even assuming a very low \aco ($\sim 1$, so far observed mostly in ULIRGs)
 the \hdue content is higher than typical values for normal galaxies of the same mass.

When considering the neutral gas still present in the disk, we find an enhanced molecular-to-neutral gas ratio $R_{mol}$ with respect to undisturbed galaxies \citep{Wang+2020} (even when using the lowest \aco), and at the same time a total gas fraction which is in good agreement with the scaling relations for normal spirals found by \citet{Catinella+2018}.

These results strongly suggest that the gas compression caused by the ram pressure in the peak stripping phase causes the conversion of large amounts of HI into the molecular phase in the disk, possibly implying that only part of the HI gets efficiently stripped.

\acknowledgments
{We acknowledge funding from the agreement ASI-INAF n.2017-14-H.0, as well as from the INAF main-stream funding programme.
B.~V., M.~G. and R.~P. also acknowledge the Italian PRIN-Miur 2017 (PI A. Cimatti).
This project has received funding from the European Research Council (ERC) under the European Union's Horizon 2020 research and innovation programme (grant agreement No. 833824, GASP project and grant agreement No. 679627, FORNAX project).
We acknowledge S. Tonnesen and Y. Jaff\'e for the stimulating discussion and  J. Wang for sharing in electronic format her data.
This paper makes use of the following ALMA data: ADS/JAO.ALMA\#2017.1.00496.S. ALMA is a partnership of ESO (representing its member states), NSF (USA) and NINS (Japan), together with NRC (Canada) and NSC and ASIAA (Taiwan), in cooperation with the Republic of Chile. The Joint ALMA Observatory is operated by ESO, AUI/NRAO and NAOJ. 
Based on observations collected by the European Organisation for Astronomical Research in the Southern Hemisphere under ESO program 196.B-0578 (VLT/MUSE).
This research made use of APLpy, an open-source plotting package for Python (Robitaille and Bressert, 2012; Robitaille, 2019).
}

\vspace{5mm}
\facilities{ALMA, VLT(MUSE)}
\software{
          CASA \citep{CASA}
          }



\begin{thebibliography}{}
\expandafter\ifx\csname natexlab\endcsname\relax\def\natexlab#1{#1}\fi
\providecommand{\url}[1]{\href{#1}{#1}}
\providecommand{\dodoi}[1]{doi:~\href{http://doi.org/#1}{\nolinkurl{#1}}}
\providecommand{\doeprint}[1]{\href{http://ascl.net/#1}{\nolinkurl{http://ascl.net/#1}}}
\providecommand{\doarXiv}[1]{\href{https://arxiv.org/abs/#1}{\nolinkurl{https://arxiv.org/abs/#1}}}

\bibitem[{Accurso {et~al.}(2017)Accurso, Saintonge, Catinella, Cortese,
  Dav{\'{e}}, Dunsheath, Genzel, Gracia-Carpio, Heckman, Jimmy, Kramer, Li,
  Lutz, Schiminovich, Schuster, Sternberg, Sturm, Tacconi, Tran, \&
  Wang}]{Accurso2017}
Accurso, G., Saintonge, A., Catinella, B., {et~al.} 2017, Monthly Notices of
  the Royal Astronomical Society, 470, 4750, \dodoi{10.1093/mnras/stx1556}

\bibitem[{Amorin {et~al.}(2016)Amorin, Mu{\~{n}}oz-Tu{\~{n}}{\'{o}}n, Aguerri,
  \& Planesas}]{AMorin+2016}
Amorin, R., Mu{\~{n}}oz-Tu{\~{n}}{\'{o}}n, C., Aguerri, J.~A.~L., \& Planesas,
  P. 2016, {\textbackslash}aap, 588, A23, \dodoi{10.1051/0004-6361/201526397}

\bibitem[{Bellhouse {et~al.}(2017)Bellhouse, Jaff{\'{e}}, Hau, McGee,
  Poggianti, Moretti, Gullieuszik, Bettoni, Fasano, D’Onofrio, Fritz,
  Omizzolo, Sheen, \& Vulcani}]{gaspII}
Bellhouse, C., Jaff{\'{e}}, Y.~L., Hau, G. K.~T., {et~al.} 2017, The
  Astrophysical Journal, 844, 49, \dodoi{10.3847/1538-4357/aa7875}

\bibitem[{Bigiel {et~al.}(2008)Bigiel, Leroy, Walter, Brinks, de~Blok, Madore,
  \& Thornley}]{Bigiel2008}
Bigiel, F., Leroy, A., Walter, F., {et~al.} 2008, The Astronomical Journal,
  136, 2846, \dodoi{10.1088/0004-6256/136/6/2846}

\bibitem[{Bigiel {et~al.}(2010)Bigiel, Walter, Blitz, Brinks, de~Blok, Madore,
  \& Madore}]{Bigiel2010}
Bigiel, F., Walter, F., Blitz, L., {et~al.} 2010, The Astronomical Journal,
  140, 1194, \dodoi{10.1088/0004-6256/140/5/1194}

\bibitem[{Bigiel {et~al.}(2011)Bigiel, Leroy, Walter, Brinks, de~Blok, Kramer,
  Rix, Schruba, Schuster, Usero, \& Wiesemeyer}]{Bigiel2011}
Bigiel, F., Leroy, A.~K., Walter, F., {et~al.} 2011, The Astrophysical Journal,
  730, L13, \dodoi{10.1088/2041-8205/730/2/L13}

\bibitem[{Bolatto {et~al.}(2013)Bolatto, Wolfire, \& Leroy}]{Bolatto2013}
Bolatto, A.~D., Wolfire, M., \& Leroy, A.~K. 2013, Annual Review of Astronomy
  and Astrophysics, vol. 51, issue 1, pp. 207-268, 51, 207,
  \dodoi{10.1146/annurev-astro-082812-140944}

\bibitem[{Bolatto {et~al.}(2017)Bolatto, Wong, Utomo, Blitz, Vogel,
  S{\'{a}}nchez, Barrera-Ballesteros, Cao, Colombo, Dannerbauer,
  Garc{\'{i}}a-Benito, Herrera-Camus, Husemann, Kalinova, Leroy, Leung, Levy,
  Mast, Ostriker, Rosolowsky, Sandstrom, Teuben, van~de Ven, \&
  Walter}]{Bolatto+2017}
Bolatto, A.~D., Wong, T., Utomo, D., {et~al.} 2017, The Astrophysical Journal,
  846, 159, \dodoi{10.3847/1538-4357/aa86aa}

\bibitem[{Boselli {et~al.}(2014)Boselli, Cortese, Boquien, Boissier, Catinella,
  Gavazzi, Lagos, \& Saintonge}]{Boselli+2014}
Boselli, A., Cortese, L., Boquien, M., {et~al.} 2014, Astronomy {\&}
  Astrophysics, 564, A67, \dodoi{10.1051/0004-6361/201322313}

\bibitem[{Boselli {et~al.}(1997)Boselli, Gavazzi, Lequeux, Buat, Casoli,
  Dickey, \& Donas}]{Boselli+1997}
Boselli, A., Gavazzi, G., Lequeux, J., {et~al.} 1997, Astronomy and
  Astrophysics, 327, 522.
\newblock \url{http://adsabs.harvard.edu/abs/1997A%26A...327..522B}

\bibitem[{Casasola {et~al.}(2004)Casasola, Bettoni, \&
  Galletta}]{Casasola+2004}
Casasola, V., Bettoni, D., \& Galletta, G. 2004, Astronomy {\&} Astrophysics,
  422, 941, \dodoi{10.1051/0004-6361:20040283}

\bibitem[{Casasola {et~al.}(2020)Casasola, Bianchi, De~Vis, Magrini, Corbelli,
  Clark, Fritz, Nersesian, Viaene, Baes, Cassar{\`{a}}, Davies, De~Looze,
  Dobbels, Galametz, Galliano, Jones, Madden, Mosenkov, Tr{\v{c}}ka, \&
  Xilouris}]{Casasola+2020}
Casasola, V., Bianchi, S., De~Vis, P., {et~al.} 2020, {\textbackslash}aap, 633,
  A100, \dodoi{10.1051/0004-6361/201936665}

\bibitem[{Catinella {et~al.}(2018)Catinella, Saintonge, Janowiecki, Cortese,
  Dav{\'{e}}, Lemonias, Cooper, Schiminovich, Hummels, Fabello, Ger{\'{e}}b,
  Kilborn, \& Wang}]{Catinella+2018}
Catinella, B., Saintonge, A., Janowiecki, S., {et~al.} 2018, Monthly Notices of
  the Royal Astronomical Society, 476, 875, \dodoi{10.1093/mnras/sty089}

\bibitem[{Chabrier(2003)}]{Chabrier2003}
Chabrier, G. 2003, The Publications of the Astronomical Society of the Pacific,
  Volume 115, Issue 809, pp. 763-795., 115, 763, \dodoi{10.1086/376392}

\bibitem[{Corbelli {et~al.}(2012)Corbelli, Bianchi, Cortese, Giovanardi,
  Magrini, Pappalardo, Boselli, Bendo, Davies, Grossi, Madden, Smith, Vlahakis,
  Auld, Baes, De~Looze, Fritz, Pohlen, \& Verstappen}]{Corbelli+2012}
Corbelli, E., Bianchi, S., Cortese, L., {et~al.} 2012, Astronomy {\&}
  Astrophysics, 542, A32, \dodoi{10.1051/0004-6361/201117329}

\bibitem[{Cortese {et~al.}(2011)Cortese, Catinella, Boissier, Boselli, \&
  Heinis}]{Cortese+2011}
Cortese, L., Catinella, B., Boissier, S., Boselli, A., \& Heinis, S. 2011,
  Monthly Notices of the Royal Astronomical Society, 415, 1797,
  \dodoi{10.1111/j.1365-2966.2011.18822.x}

\bibitem[{Cortese {et~al.}(2017)Cortese, Catinella, \&
  Janowiecki}]{Cortese+2017}
Cortese, L., Catinella, B., \& Janowiecki, S. 2017, The Astrophysical Journal,
  848, L7, \dodoi{10.3847/2041-8213/aa8cc3}

\bibitem[{Deb {et~al.}(2020)Deb, Verheijen, Gullieuszik, Poggianti, van Gorkom,
  Ramatsoku, Serra, Moretti, Vulcani, Bettoni, Jaff{\'{e}}, Tonnesen, \&
  Fritz}]{Deb+2020}
Deb, T., Verheijen, M. A.~W., Gullieuszik, M., {et~al.} 2020,
  {\textbackslash}mnras, \dodoi{10.1093/mnras/staa968}

\bibitem[{Fern{\'{a}}ndez {et~al.}(2016)Fern{\'{a}}ndez, Gim, Gorkom, Yun,
  Momjian, Popping, Chomiuk, Hess, Hunt, Kreckel, Lucero, Maddox, Oosterloo,
  Pisano, Verheijen, Hales, Chung, Dodson, Golap, Gross, Henning, Hibbard,
  Jaff{\'{e}}, Meyer, Meyer, Sanchez-Barrantes, Schiminovich, Wicenec, Wilcots,
  Bershady, Scoville, Strader, Tremou, Salinas, \& Ch{\'{a}}vez}]{chiles}
Fern{\'{a}}ndez, X., Gim, H.~B., Gorkom, J. H.~v., {et~al.} 2016, The
  Astrophysical Journal Letters, 824, L1, \dodoi{10.3847/2041-8205/824/1/L1}

\bibitem[{Franchetto {et~al.}(2020)Franchetto, Vulcani, Poggianti, Gullieuszik,
  Mingozzi, Moretti, Tomi{\v{c}}i{\'{c}}, Fritz, Bettoni, \&
  Jaff{\'{e}}}]{Franchetto_2020}
Franchetto, A., Vulcani, B., Poggianti, B.~M., {et~al.} 2020, The Astrophysical
  Journal, 895, 106, \dodoi{10.3847/1538-4357/ab8db9}

\bibitem[{Genzel {et~al.}(2012)Genzel, Tacconi, Combes, Bolatto, Neri,
  Sternberg, Cooper, Bouch{\'{e}}, Bournaud, Burkert, Comerford, Cox, Davis,
  Schreiber, Garcia-Burillo, Gracia-Carpio, Lutz, Naab, Newman, Saintonge,
  Shapiro, Shapley, \& Weiner}]{Genzel+2012}
Genzel, R., Tacconi, L.~J., Combes, F., {et~al.} 2012, The Astrophysical
  Journal, 746, 69, \dodoi{10.1088/0004-637X/746/1/69}

\bibitem[{George {et~al.}(2018)George, Poggianti, Gullieuszik, Fasano,
  Bellhouse, Postma, Moretti, Jaff{\'{e}}, Vulcani, Bettoni, Fritz,
  C{\^{o}}t{\'{e}}, Ghosh, Hutchings, Mohan, Sreekumar, Stalin, Subramaniam, \&
  Tandon}]{George+2018}
George, K., Poggianti, B.~M., Gullieuszik, M., {et~al.} 2018, Monthly Notices
  of the Royal Astronomical Society, 479, 4126, \dodoi{10.1093/mnras/sty1452}

\bibitem[{Gullieuszik {et~al.}(2017)Gullieuszik, Poggianti, Moretti, Fritz,
  Jaff{\'{e}}, Hau, Bischko, Bellhouse, Bettoni, Fasano, Vulcani, D’Onofrio,
  \& Biviano}]{gaspIV}
Gullieuszik, M., Poggianti, B.~M., Moretti, A., {et~al.} 2017, The
  Astrophysical Journal, 846, 27, \dodoi{10.3847/1538-4357/aa8322}

\bibitem[{Israel {et~al.}(2015)Israel, Rosenberg, \& van~der
  Werf}]{Israel+2015}
Israel, F.~P., Rosenberg, M.~J.~F., \& van~der Werf, P. 2015,
  {\textbackslash}aap, 578, A95, \dodoi{10.1051/0004-6361/201425175}

\bibitem[{J{\'{a}}chym {et~al.}(2014)J{\'{a}}chym, Combes, Cortese, Sun, \&
  Kenney}]{Jachym2014}
J{\'{a}}chym, P., Combes, F., Cortese, L., Sun, M., \& Kenney, J. D.~P. 2014,
  The Astrophysical Journal, 792, 11, \dodoi{10.1088/0004-637X/792/1/11}

\bibitem[{Jachym {et~al.}(2017)Jachym, Sun, Kenney, Cortese, Combes, Yagi,
  Yoshida, Palous, \& Roediger}]{Jachym2017}
Jachym, P., Sun, M., Kenney, J. D.~P., {et~al.} 2017, The Astrophysical
  Journal, Volume 839, Issue 2, article id. 114, 15 pp. (2017)., 839,
  \dodoi{10.3847/1538-4357/aa6af5}

\bibitem[{J{\'{a}}chym {et~al.}(2019)J{\'{a}}chym, Kenney, Sun, Combes,
  Cortese, Scott, Sivanandam, Brinks, Roediger, Palou{\v{s}}, \&
  Fumagalli}]{Jachym+2019}
J{\'{a}}chym, P., Kenney, J. D.~P., Sun, M., {et~al.} 2019, The Astrophysical
  Journal, 883, 145, \dodoi{10.3847/1538-4357/ab3e6c}

\bibitem[{Kenney \& Young(1989)}]{Kenney+1989}
Kenney, J. D.~P., \& Young, J.~S. 1989, The Astrophysical Journal, 344, 171,
  \dodoi{10.1086/167787}

\bibitem[{Kennicutt(1998)}]{Kennicutt1998}
Kennicutt, R.~C. 1998, Annual Review of Astronomy and Astrophysics, 36, 189,
  \dodoi{10.1146/annurev.astro.36.1.189}

\bibitem[{Leroy {et~al.}(2008)Leroy, Walter, Brinks, Bigiel, de~Blok, Madore,
  \& Thornley}]{Leroy2008}
Leroy, A.~K., Walter, F., Brinks, E., {et~al.} 2008, The Astronomical Journal,
  136, 2782, \dodoi{10.1088/0004-6256/136/6/2782}

\bibitem[{Leroy {et~al.}(2013)Leroy, Walter, Sandstrom, Schruba, Munoz-Mateos,
  Bigiel, Bolatto, Brinks, de~Blok, Meidt, Rix, Rosolowsky, Schinnerer,
  Schuster, \& Usero}]{Leroy2013}
Leroy, A.~K., Walter, F., Sandstrom, K., {et~al.} 2013, The Astronomical
  Journal, 146, 19, \dodoi{10.1088/0004-6256/146/2/19}

\bibitem[{McMullin(2007)}]{CASA}
McMullin, J.P.;~Waters, B. S. D. Y. W. G.~K. 2007, in Astronomical Data
  Analysis Software and Systems XVI, ed. F.~H. Richard A.~Shaw \& D.~J. Bell
  (ASP Conf. Series, 376), 127.
\newblock \url{http://aspbooks.org/custom/publications/paper/376-0127.html}

\bibitem[{Moretti {et~al.}(2018)Moretti, Paladino, Poggianti, D’Onofrio,
  Bettoni, Gullieuszik, Jaff{\'{e}}, Vulcani, Fasano, Fritz, \&
  Torstensson}]{gaspX}
Moretti, A., Paladino, R., Poggianti, B.~M., {et~al.} 2018, Monthly Notices of
  the Royal Astronomical Society, 480, 2508, \dodoi{10.1093/mnras/sty2021}

\bibitem[{Moretti {et~al.}(2020)Moretti, Paladino, Poggianti, Serra, Roediger,
  Gullieuszik, Tomi{\v{c}}i{\'{c}}, Radovich, Vulcani, Jaff{\'{e}}, Fritz,
  Bettoni, Ramatsoku, \& Wolter}]{GASPXXII}
---. 2020, The Astrophysical Journal, 889, 9, \dodoi{10.3847/1538-4357/ab616a}

\bibitem[{Poggianti {et~al.}(2017{\natexlab{a}})Poggianti, Moretti,
  Gullieuszik, Fritz, Jaff{\'{e}}, Bettoni, Fasano, Bellhouse, Hau, Vulcani,
  Biviano, Omizzolo, Paccagnella, D’Onofrio, Cava, Sheen, Couch, \&
  Owers}]{gaspI}
Poggianti, B.~M., Moretti, A., Gullieuszik, M., {et~al.} 2017{\natexlab{a}},
  The Astrophysical Journal, 844, 48, \dodoi{10.3847/1538-4357/aa78ed}

\bibitem[{Poggianti {et~al.}(2017{\natexlab{b}})Poggianti, Jaff{\'{e}},
  Moretti, Gullieuszik, Radovich, Tonnesen, Fritz, Bettoni, Vulcani, Fasano,
  Bellhouse, Hau, \& Omizzolo}]{poggianti2017}
Poggianti, B.~M., Jaff{\'{e}}, Y.~L., Moretti, A., {et~al.} 2017{\natexlab{b}},
  Nature, 548, 304, \dodoi{10.1038/nature23462}

\bibitem[{Poggianti {et~al.}(2019)Poggianti, Ignesti, Gitti, Wolter, Brighenti,
  Biviano, George, Vulcani, Gullieuszik, Moretti, Paladino, Bettoni,
  Franchetto, Jaff{\'{e}}, Radovich, Roediger, Tomi{\v{c}}i{\'{c}}, Tonnesen,
  Bellhouse, Fritz, \& Omizzolo}]{GASPXXIII}
Poggianti, B.~M., Ignesti, A., Gitti, M., {et~al.} 2019, {\textbackslash}apj,
  887, 155, \dodoi{10.3847/1538-4357/ab5224}

\bibitem[{Ramatsoku {et~al.}(2020)Ramatsoku, Serra, Poggianti, Moretti,
  Gullieuszik, Bettoni, Deb, van Gorkom, Jaff{\`{e}}, Tonnesen, Verheijen,
  Vulcani, \& de~Blok}]{GASPXXVIII}
Ramatsoku, M, A., Serra, P., Poggianti, B.~M., {et~al.} 2020, A
  {\textbackslash}{\&} A submitted

\bibitem[{Ramatsoku {et~al.}(2019)Ramatsoku, Serra, Poggianti, Moretti,
  Gullieuszik, Bettoni, Deb, Fritz, van Gorkom, Jaff{\'{e}}, Tonnesen,
  Verheijen, Vulcani, Hugo, J{\'{o}}zsa, Maccagni, Makhathini, Ramaila,
  Smirnov, \& Thorat}]{gaspXVII}
Ramatsoku, M., Serra, P., Poggianti, B.~M., {et~al.} 2019, Monthly Notices of
  the Royal Astronomical Society, 487, 4580, \dodoi{10.1093/mnras/stz1609}

\bibitem[{Saintonge {et~al.}(2011{\natexlab{a}})Saintonge, Kauffmann, Kramer,
  Tacconi, Buchbender, Catinella, Fabello, Graci{\'{a}}-Carpio, Wang, Cortese,
  Fu, Genzel, Giovanelli, Guo, Haynes, Heckman, Krumholz, Lemonias, Li, Moran,
  Rodriguez-Fernandez, Schiminovich, Schuster, \& Sievers}]{Saintonge2011}
Saintonge, A., Kauffmann, G., Kramer, C., {et~al.} 2011{\natexlab{a}}, Monthly
  Notices of the Royal Astronomical Society, 415, 32,
  \dodoi{10.1111/j.1365-2966.2011.18677.x}

\bibitem[{Saintonge {et~al.}(2011{\natexlab{b}})Saintonge, Kauffmann, Wang,
  Kramer, Tacconi, Buchbender, Catinella, Graci{\'{a}}-Carpio, Cortese,
  Fabello, Fu, Genzel, Giovanelli, Guo, Haynes, Heckman, Krumholz, Lemonias,
  Li, Moran, Rodriguez-Fernandez, Schiminovich, Schuster, \&
  Sievers}]{Saintonge2011b}
Saintonge, A., Kauffmann, G., Wang, J., {et~al.} 2011{\natexlab{b}}, Monthly
  Notices of the Royal Astronomical Society, 415, 61,
  \dodoi{10.1111/j.1365-2966.2011.18823.x}

\bibitem[{Saintonge {et~al.}(2017)Saintonge, Catinella, Tacconi, Kauffmann,
  Genzel, Cortese, Dav{\'{e}}, Fletcher, Graci{\'{a}}-Carpio, Kramer, Heckman,
  Janowiecki, Lutz, Rosario, Schiminovich, Schuster, Wang, Wuyts, Borthakur,
  Lamperti, \& Roberts-Borsani}]{Saintonge2017}
Saintonge, A., Catinella, B., Tacconi, L.~J., {et~al.} 2017, The Astrophysical
  Journal Supplement Series, 233, 22, \dodoi{10.3847/1538-4365/aa97e0}

\bibitem[{Sandstrom {et~al.}(2013)Sandstrom, Leroy, Walter, Bolatto, Croxall,
  Draine, Wilson, Wolfire, Calzetti, Kennicutt, Aniano, Donovan~Meyer, Usero,
  Bigiel, Brinks, de~Blok, Crocker, Dale, Engelbracht, Galametz, Groves, Hunt,
  Koda, Kreckel, Linz, Meidt, Pellegrini, Rix, Roussel, Schinnerer, Schruba,
  Schuster, Skibba, van~der Laan, Appleton, Armus, Brandl, Gordon, Hinz,
  Krause, Montiel, Sauvage, Schmiedeke, Smith, \& Vigroux}]{Sandstrom+2013}
Sandstrom, K.~M., Leroy, A.~K., Walter, F., {et~al.} 2013, The Astrophysical
  Journal, 777, 5, \dodoi{10.1088/0004-637X/777/1/5}

\bibitem[{Schmidt(1959)}]{Schmidt1959}
Schmidt, M. 1959, The Astrophysical Journal, 129, 243, \dodoi{10.1086/146614}

\bibitem[{Schruba {et~al.}(2012)Schruba, Leroy, Walter, Bigiel, Brinks,
  de~Blok, Kramer, Rosolowsky, Sandstrom, Schuster, Usero, Weiss, \&
  Wiesemeyer}]{Schruba+2012}
Schruba, A., Leroy, A.~K., Walter, F., {et~al.} 2012, {\textbackslash}aj, 143,
  138, \dodoi{10.1088/0004-6256/143/6/138}

\bibitem[{Serra {et~al.}(2015)Serra, Westmeier, Giese, Jurek, Fl{\"{o}}er,
  Popping, Winkel, Van~der Hulst, Meyer, Koribalski, Staveley-Smith, \&
  Courtois}]{sofia}
Serra, P., Westmeier, T., Giese, N., {et~al.} 2015, Monthly Notices of the
  Royal Astronomical Society, \dodoi{10.1093/mnras/stv079}

\bibitem[{Vulcani {et~al.}(2018)Vulcani, Poggianti, Gullieuszik, Moretti,
  Tonnesen, Jaff{\'{e}}, Fritz, Fasano, \& Bettoni}]{vulcani+2018_sf}
Vulcani, B., Poggianti, B.~M., Gullieuszik, M., {et~al.} 2018, The
  Astrophysical Journal, 866, L25, \dodoi{10.3847/2041-8213/aae68b}

\bibitem[{Wang {et~al.}(2020)Wang, Catinella, Saintonge, Pan, Serra, \&
  Shao}]{Wang+2020}
Wang, J., Catinella, B., Saintonge, A., {et~al.} 2020, The Astrophysical
  Journal, 890, 63, \dodoi{10.3847/1538-4357/ab68dd}

\bibitem[{Watson \& Koda(2017)}]{WatsonKoda2016}
Watson, L.~C., \& Koda, J. 2017, in Outskirts of Galaxies, ed. L.~J.~C. Knapen,
  J.~H. \& G.~de~Paz~A. (Astrophysics and Space Science Proceedings), 175--207.
\newblock \url{http://arxiv.org/abs/1612.05275}

\bibitem[{Wolfire {et~al.}(2010)Wolfire, Hollenbach, \& McKee}]{Wolfire+2010}
Wolfire, M.~G., Hollenbach, D., \& McKee, C.~F. 2010, The Astrophysical
  Journal, 716, 1191, \dodoi{10.1088/0004-637X/716/2/1191}

\end{thebibliography}
\end{document}